\documentclass[12pt]{article}

\begin{document}

\author{A.I.Volokitin$^{1,2}$ , B.N.J.Persson$^1$ and H.Ueba$^3$ \\
\\
$^1$Institut f\"ur Festk\"orperforschung, Forschungszentrum \\
J\"ulich, D-52425, Germany\\
$^2$Samara State Technical University, 443100 Samara,\\
Russia\\
$^3$Department of Electronics, Toyama University, Gofuku,\\
\thinspace \thinspace \thinspace \thinspace Toyama, 930-8555, Japan}
\title{Enhancement of the non-contact friction between closely spaced bodies by
two-dimension systems}
\maketitle

\begin{abstract}
We consider the effect of an external bias voltage and the spatial variation
of the surface potential, on the damping of cantilever vibrations. The
electrostatic friction is due to energy losses in the sample created by the
electromagnetic field from the oscillating charges induced on the surface of
the tip by the bias voltage and spatial variation of the surface potential.
A similar effect arises when the tip is oscillating in the electrostatic
field created by charged defects in a dielectric substrate. The
electrostatic friction is compared with the van der Waals friction
originating from the fluctuating electromagnetic field due to quantum and
thermal fluctuation of the current density inside the bodies. We show that
the electrostatic and van der Waals friction can be greatly enhanced if on
the surfaces of the sample and the tip there are two-dimension (2D) systems,
e.g. a 2D-electron system or incommensurate layers of adsorbed ions
exhibiting acoustic vibrations. We show that the damping of the cantilever
vibrations due to the electrostatic friction may be of similar magnitude as
the damping observed in recent experiments of Stipe \textit{et al}
[B.C.Stipe, H.J.Mamin, T.D.Stowe, T.W.Kenny, and D.Rugar, Phys.Rev. Lett.%
\textbf{87}, 0982001]. We also show that at short separation the van der
Waals friction may be large enough to be measured experimentally.
\end{abstract}

\section{ Introduction}

A great deal of attention has been devoted to non-contact friction between
an atomic force microscope tip and a substrate \cite
{Dorofeev,Gostmann,Stipe,Mamin,Hoffmann}. This problem is related to the
role of non-contact friction for ultrasensitive force detection experiments.
The ability to detect small forces is inextricably linked to friction via
the fluctuation-dissipation theorem. According to this theorem, the random
force that make a small particle jitter would also cause friction if the
particle were dragged through the medium. For example, the detection of
single spins by magnetic resonance force microscopy \cite{Rugar}, which has
been proposed for three-dimensional atomic imaging \cite{Sidles} and quantum
computation \cite{Berman}, will require force fluctuations (and consequently
the friction) to be reduced to unprecedented levels. In addition, the search
for quantum gravitation effects at short length scale \cite{Arkani}, and
future measurements of the dynamical Casimir forces \cite{Mohideen}, may
eventually be limited by non-contact friction effects. Non-contact friction
is also responsible for the frictional drag force between two-dimensional
(2D) quantum wells \cite{Gramila1,Gramila2,Sivan}.

In non-contact friction the bodies are separated by a potential barrier
thick enough to prevent electrons or other particles with a finite rest mass
from tunneling across it, but allowing interaction via the long-range
electromagnetic field, which is always present in the gap between bodies and
can have different origin. The presence of an inhomogeneous tip-sample
electric fields is difficult to avoid, even under the best experimental
conditions \cite{Stipe}. For example, even if both the tip and the sample
were metallic single crystals, the tip would still have corners, and more
than one crystallographic plane exposed. The presence of atomic steps,
adsorbates, and other defects will also contribute to the spatial variation
of the surface potential. This is referred to as ``patch effect''. The
surface potential can also be easily changed by applying a voltage between
the tip and the sample. An inhomogeneous electric field can also be created
by charged defects embedded in a dielectric sample. The relative motion of
the charged bodies will produce friction which will be denoted as the
\textit{electrostatic friction}.

The electromagnetic field can also be created by the fluctuating current
density, due to thermal and quantum fluctuations inside the solids. This
fluctuating electromagnetic field gives rise to the well-known long-range
attractive van der Waals interaction between two bodies \cite{Lifshitz1},
and is responsible for radiative heat transfer. If the bodies are in
relative motion, the same fluctuating electromagnetic field will give rise
to a friction which is frequently named as the van der Waals friction.

Recently Stipe \textit{et.al.}\cite{Stipe} observed non-contact friction
between a gold surface and a gold-coated cantilever as a function of
tip-sample spacing $d$, temperature $T$, and bias voltage $V$. The friction
force $F$ acting on the tip was found to be proportional to the velocity $v$%
, $F=\Gamma v$. For vibration of the tip parallel to the surface they found $%
\Gamma (d)=\alpha (T)(V^2+V_0^2)/d^n$, where $n=1.3\pm 0.2,$ and $V_0\sim
0.2\,\mathrm{V.}$ At 295\textrm{K, }for the spacing $d=$ 100\textrm{\AA\ }%
they found $\Gamma =1.5\times 10^{-13}\mathrm{\,kgs}^{-1}$. An applied
voltage of 1 V resulted in a friction $\Gamma =3\times 10^{-12}$kg/s at 300
K with $d=20$nm.

In Ref.\cite{Stipe} the non-contact friction has also measured for fused
silica samples. Near the silica surface the friction was found to be an
order of magnitude larger than for the gold sample. The silica sample had
been irradiated with $\gamma $ rays which produce $E^{\prime }$ centers (Si
dangling bonds) at a density of $7\times 10^{17}$cm$^{-3}$. Although the
sample is electrically neutral overall, the $E^{\prime }$ centers are known
to be positively charged, creating enhanced field inhomogeneity and causing
the non-contact friction to rise another order of magnitude.

Attempts to explain the observed friction in terms of the van der Waals
friction have not met with much success since the van der Waals friction for
good conductors like copper has been shown \cite
{Volokitin6,Volokitin7,Persson and Volokitin} to be many orders of magnitude
smaller than the friction observed by Stipe \textit{et.al.}. In \cite
{Greffet2} it was proposed that the van der Waals friction may be strongly
enhanced between a high resistivity mica substrate and silica tip. However
in \cite{Stipe} the mica substrate and silica tip were coated by gold films
thick enough to completely screen the electrodynamic interaction between the
underlying dielectrics.

At small separation $d \sim 1$nm, resonant photon tunneling between
adsorbate vibrational modes on the tip and the sample may increase the
friction by seven order of magnitude in comparison with the good conductors
surfaces \cite{Volokitin4,Volokitin5}. However, the distance dependence ($%
\sim 1/d^{6}$) is stronger than observed experimentally \cite{Stipe}.

Recently, a theory of noncontact friction was suggested where the friction
arises from Ohmic losses associated with the electromagnetic field created
by moving charges induced by the bias voltage \cite{Chumak}. In the case of
a spherical tip this theory predict the same weak distance dependence of the
friction as observed in the experiment, but the magnitude of the friction is
many orders of magnitude smaller than found experimentally. However, we have
shown that the electrostatic friction can be greatly enhanced if there is an
incommensurate adsorbed layer exhibiting acoustic vibrations \cite
{Volokitin8}. This theory gives an explanation for the experimentally
observed bias voltage contribution to the non-contact friction.

In this article we extend the theory presented in
\cite{Volokitin8} to include the contribution to friction from the
spatial variation of the surface potential and from the spatial
fluctuation of the electric charge of charged defects in the bulk
of the dielectric. We also show that the electrostatic friction as
well as the van der Waals friction can be greatly enhanced for
2D-systems, e.g. a 2D-electron system or an incommensurate layer
of adsorbed ions exhibiting acoustic vibrations. The origin of
this enhancement is related to the fact that the screening in
2D-systems is much less effective than for 3D-systems. An atomic
force microscope tip charged by the bias voltage, or by the
spatial variation of the surface potential, and moving close to
the metal surface will induce ``image'' charge in the 2D-system.
Because of the finite response time this ``image'' charge will lag
behind the tip, and this effect result in force acting on the tip,
referred to as the ``electrostatic friction''. However, the weaker
screening effect in the 2D-system will result in a much weaker
restoring force, which occurs when the ``image charge'' is
displaced from the equilibrium position, and this result in larger
lag of the ``image'' charge in 2D-systems in comparison with
3D-systems.

Another contribution to the friction from the electric field, is
associated with the time-dependent stress acting on the surface of
the surface due to the tip oscillations. This stress can excite
acoustic phonons, or induce non-adiabatic time-dependent
deformation. In this article we develop theories of phonon and
internal friction due to  the time-dependent stress acting on the
surface. We show that  this stress depends on the bias voltage as
$V^2\,$ resulting in to the friction coefficient $\Gamma \sim
V^4$. Thus this mechanism can be ruled out as an explanation of
the experimental data observed in \cite{Stipe}, where $\Gamma \sim
V^2.\,$ In the case of phonon friction only phonons with $q<\omega
/c_s$ can be excited, where $q$ is the component of the
wave-vector parallel to the surface of the substrate, $\omega $ is
the frequency of the tip oscillations, and $c_s$ is the sound
velocity. Thus in the phase space the area occupied by the excited
phonons $\sim (\omega /c_s)^2$. For electromagnetic mechanisms of
the friction (which include the electrostatic and van der Waals
friction) all components of the electromagnetic field with $
q<d_1^{-1}, $ where $d_1$ is the radius of interaction (see below,
typically $d_1\sim 100 $nm), are important. Thus for the metal
substrate in the typical case $(\omega d_1/c_s)^2\ll 1$, the
phonon friction is negligible in the comparison with the
electromagnetic friction.

\section{Electrostatic friction due to a bias voltage and the spatial
variation of the surface potential}

\subsection{A general theory}

We begin by considering a model in which the tip of a metallic
cantilever of length $L$ is a section of a cylindrical surface
with the radius of curvature $R$ (Fig.1). The cantilever is
perpendicular to a flat sample surface, which occupies the $xy$
plane, with the $z$- axis pointing outside the sample. The tip
displacement $\mathbf{u}(t)=\hat xu_0e^{-i\omega t}$ is assumed to
be parallel to the surface (along the $x$ axis), which will be a
good approximation when the oscillation amplitudes $u_0$ is
sufficiently small. The cantilever width $w$, i.e. the size in the
direction perpendicular to the $xz$ plane, is taken to be much
larger than the thickness $c$ ($w\gg c$), and $d$ is the
separation between the tip and the sample surface. It is
straightforward to obtain the static electric field distribution
in the practically important case os small distances $d$ such that
the electrostatic field of the entire cylinder effectively the
same as that due to its bottom part. ( The criterion that $d$ ut
satisfy for this to be the case is given by $\sqrt{d/R}\ll 1\,.$)
The problem is then reduced to solving the two-dimension Laplace
equation with the boundary conditions that the potential has
constant values $V\,$ and $0\,$at the metallic surfaces of the tip
and the substrate. The electric field distribution outside the
conductors is equal to the field due to two charged wires passing
through points at $z=\pm =\pm \sqrt{(d+R)^2-R^2}$ \cite{Landau1}.
The wires have charges $\pm Q$ per unit length, $Q=CV$, where
$C^{-1}=2\ln [(d+R+d_1)/R]$. The electric potential at a point
$\mathbf{r}$ exterior to the tip and sample is given by
\begin{eqnarray}
\varphi _0(\mathbf{r)} &&\mathbf{=}-2Q\left[ \ln |\mathbf{r-r}_{+}|-\ln |%
\mathbf{r-r}_{-}|\right]  \nonumber  \label{one.one} \\
&=&Q\int_{-\infty }^\infty \frac{dq}{|q|}e^{iqx}\left[
e^{-|q||z-z_{+}|}-e^{-|q||z-z_{-}|}\right] .  \label{one.one}
\end{eqnarray}
where $\mathbf{r}_{\pm }=\pm \hat zd_1$. The attractive cantilever-surface
force can be calculated straightforwardly using Eq.(\ref{one.one}) \cite
{Chumak}.

A somewhat different picture applies in the case of a oscillating charged
tip. The cantilever charge is not changed when its tip moves parallel to the
surface, while the sample charge varies in time at any fixed point. The
electric field from the oscillating tip will induce an electric charge in
the sample and this will result in to induced electric field outside the
sample. The oscillating electric potential due to the tip oscillation at a
point $\mathbf{r}$ exterior to the tip and sample is given by
\begin{equation}
\varphi _1(\mathbf{r,t)=}\varphi _1(\mathbf{r)e}^{-i\omega t}+c.c.,
\label{one.two}
\end{equation}
where
\begin{equation}
\varphi _1(\mathbf{r)=}iQu_0\int_{-\infty }^\infty \frac{dqq}{|q|}%
e^{iqx}\left[ e^{-|q||z-z_{+}|}-e^{-|q||z-z_{-}|}R_p(q,\omega )\right] ,
\label{one.three}
\end{equation}
and $R_p(q,\omega )$ is the reflection amplitude for the $p-$ polarized
electromagnetic waves. The electric field is given by $\mathbf{E}(\mathbf{r}%
)=-\mathbf{\nabla }\varphi (\mathbf{r})$. The energy dissipation
per unit time induced by the electromagnetic field inside of the
metallic substrate is determined by integrating the Poynting
vector over the surface of the metal, and is given by
\begin{eqnarray}
P &=&\frac c{4\pi }\int dS\hat z\cdot [\mathbf{E(r})\mathbf{\times B}^{*}%
\mathbf{(r})]_{z=+0}+c.c.=-\frac{i\omega }{4\pi }\int dS\left( \varphi _1(%
\mathbf{r)}\frac d{dz}\varphi _1^{*}(\mathbf{r)}\right) _{z=+0}+c.c.
\nonumber \\
\  &=&4\omega Q^2|u_0|^2w\int_0^\infty dqqe^{-2qd_1}\mathrm{Im}R_p(\omega ,q)
\label{one.four}
\end{eqnarray}
Taking into account that the energy dissipation per unit time must be equal
to $2\omega ^2\Gamma \left| u_0\right| ^2$, using (\ref{one.four}) gives the
friction coefficient:
\begin{equation}
\Gamma =\lim_{\omega \to 0}2C^2V^2w\int_0^\infty dqqe^{-2qd_1}\frac{\mathrm{%
Im}R_p(\omega ,q)}\omega ,  \label{biasone}
\end{equation}
Without derivation Eq.(\ref{biasone}\} was firstly presented in
\cite {Volokitin8}. Now we assume that the electric potential on
the surface of the tip is inhomogeneous, consists of the domains
or ``patches''. Thus the cylinder with linear size $w$ is divided
on smaller cylinders with the linear size $w_i$: $w=\sum_iw_i\gg
w_i\gg \sqrt{dR}$, and with the surface potential $V_{is}=V+V_i$,
where $V$ is the bias voltage and $V_i$ is the randomly
fluctuating surface potential for the domain $i$. In the case of a
cylindrical tip geometry all domains give independent contribution
to
friction which can be obtained from Eq.(\ref{biasone}) after replacement $%
V\rightarrow V+V_i$ and $w\rightarrow w_i$ . The contribution to friction
from all domains is given by
\begin{eqnarray}
\Gamma  &=&\sum_i\Gamma _i=\sum_i\lim_{\omega \to
0}2C^2(V+V_i)^2w_i\int_0^\infty dqqe^{-2qd_1}\frac{\mathrm{Im}R_p(\omega ,q)}%
\omega   \nonumber  \label{biastwo} \\
&=&\lim_{\omega \to 0}2C^2(V^2+V_0^2)w\int_0^\infty dqqe^{-2qd_1}\frac{%
\mathrm{Im}R_p(\omega ,q)}\omega   \label{biastwo}
\end{eqnarray}
where we take into account that the average value of the fluctuating surface
potential $\langle V_i\rangle =\sum_iw_iV_i=0$ and $V_0^2=\sum_iw_iV_i^2/w,$
so that $V_0$ is the root mean square variation of the surface potential.
According to Eq.(\ref{biastwo}), bias voltage and patch contributions to the
friction have the same dependence on $d.$ Sukenik \textit{et al}, studied
the root mean square variation of the surface potential due to thermally
evaporated gold using the Stark effect in sodium atoms \cite{Sukenik}. The
films were partially optically transparent with a thickness of 42 nm and
heated at 120 $^{\circ }$C for several hours in vacuum. They deduced the
magnitude of the fluctuating surface potential to be $V_0=0.15$V, and showed
that the scale of the lateral variation of the surface potential is of the
order of the film thickness. The measurement of the non-contact friction
between a gold tip and the gold sample gave $V_0\sim 0.2$V \cite{Stipe} thus
confirming the prediction of the theory that this parameter is determined by
the root mean square variation of the surface potential.

Now, let us consider spherical tip (radius $R$) with the constant voltage
surface domains with the linear size $R_i$. If $R\gg R_i\gg \sqrt{dR}$ the
domain on the apex of the tip will give the main contribution to the
friction. In this case we can neglect the spatial variation of the surface
potential and the electric field induced by the bias voltage is
approximately the same as that which would be produced in the vacuum region
between two point charges $\pm Q_i=\pm C(V+V_i)$ located at
\begin{equation}
z=\pm d_1=\pm \sqrt{3Rd/2+\sqrt{(3Rd/2)^2+Rd^3+d^4}}  \label{d1}
\end{equation}
where
\begin{equation}
C=\frac{d_1^2-d^2}{2d}  \label{C}
\end{equation}
It can been shown that the electrostatic force between the tip and the metal
surface within this approximation agrees very well with the exact expression
for a sphere above a metal surface \cite{Hudlet}. The vibrations of the tip
will produce an oscillating electromagnetic field, which in the vacuum
region coincides with the electromagnetic field of an oscillating point
charge. The friction coefficient for a point charge moving parallel to the
surface due to the electromagnetic energy losses inside the sample, is
determined by \cite{Persson2}
\begin{equation}
\Gamma _{\Vert }=\lim_{\omega \to 0}\frac{Q_i^2}2\int_0^\infty
dqq^2e^{-2qd_1}\frac{\mathrm{Im}R_p(\omega ,q)}\omega  \label{biaseight}
\end{equation}
For motion normal to the surface, $\Gamma _{\perp }=2\Gamma _{\Vert }$.
Thus, just as for the cylindrical tip geometry, for a spherical tip the
friction depends parabolically on the bias voltage. However for a spherical
tip the parabola begins from zero in contrast to a cylindrical tip, where
the parabola begins from a finite positive value.

\subsection{Clean surface}

For a clean flat surfaces the reflection coefficient is determined by the
well-known Fresnel formula
\begin{equation}
R_p=\frac{\epsilon -1}{\epsilon +1}  \label{clean}
\end{equation}
In this case, for the tip radius $R\gg d$ and for a metal with the
dielectric function $\epsilon =1+4\pi i\sigma /\omega $, where $\sigma $ is
the conductivity, Eq.(\ref{biastwo}) gives:
\begin{equation}
\Gamma _{cl}^c=\frac{w(V^2+V_0^2)}{2^6\pi \sigma d^2}  \label{biasthree}
\end{equation}
Neglecting the contribution from the spatial variation of the surface
potential, this formula was obtained recently in \cite{Chumak}using a less
general approach. With $w=7\cdot 10^{-6}$m and $\sigma =4\cdot 10^{17}$s$%
^{-1}$ ( corresponds to gold at 300 K), and with $d=20$nm and $V=1$Volt, Eq.(%
\ref{biasthree}) gives $\Gamma =2.4\cdot 10^{-20}$kg/s which is eight orders
of magnitude smaller than the experimental value $3\cdot 10^{-12}$kg/s \cite
{Stipe}.

Assuming $R>>d,$using (\ref{biaseight}) and (\ref{clean}) gives the friction
between a spherical tip and a clean sample surface
\begin{equation}
\Gamma _{cl}^s=\frac{3^{1/2}R^{1/2}V^2}{2^7d^{3/2}\pi \sigma }
\end{equation}
This expression is only a factor 1.6 smaller that the result obtained
independently in \cite{Chumak} using a less general approach. For the same
parameters as above and at $d=20$nm, the friction for a spherical tip is two
order of magnitude smaller than for the cylindrical tip.

To get insight into possible mechanisms of the enhancement of non-contact
friction it is instructive to note that qualitatively Eq.(\ref{biasthree})
can be obtained from the following simple geometrical arguments \cite{Rugar2}%
. The vibrating tip will induce current in the sample in a volume with the
spatial dimensions $L_x$, $L_y$ and $L_z$. The instantaneous dissipated
power in the sample is given by $P\sim I^2r$, where $I$ is the current and $%
r $ is the effective resistivity. The current $I$ is proportional to the tip
velocity $v_x$, and can be written as $I\sim v_xQ_t/L_x$, where $Q_t$ is the
charge of the tip. The effective resistance $r$ can be approximated by the
macroscopic relation $r=\rho L_x/L_yL_z,$ where $\rho $ is the resistivity.
Using this simple expressions for current $I$ and resistance, and using the
relation $Q_t=C_tV_s$ (where $C_t$ is the tip-sample capacitance) for the
induced charge, the instantaneous power dissipation is
\begin{equation}
P=I^2r\sim \rho \frac{v_x^2C_t^2V_s^2}{L_xL_yL_z}.  \label{biasfour}
\end{equation}
Comparing this expression with $P=\Gamma v_x^2$ we get
\begin{equation}
\Gamma \sim \rho \frac{C_t^2V_s^2}{L_xL_yL_z}  \label{biasfive}
\end{equation}
For a cylindrical tip vibrating above the clean surface $L_y\sim w$ and $%
L_x\sim L_z\sim d_1$. For $d\ll R$ the tip-sample capacitance $C_t\sim w%
\sqrt{R/8d}$ and $d_1\sim \sqrt{2dR}$. Substituting these expressions in Eq.(%
\ref{biasfive}) gives Eq.(\ref{biasthree}) to within a numerical factor of
order of unity. From Eq.(\ref{biasfive}) it follows that the friction will
increase when the thickness $L_z$ of ``dissipation volume'' decreases. This
is the reason for why 2D-systems may exhibit higher friction than 3D-systems.

\subsection{Film on-top of a high-resistivity substrate}

>From the qualitative arguments given above it follows that for a thin metal
film on-top of a high resistivity substrate, e.g. a dielectric or a high
resistivity metal, the friction will be larger, than for an infinitely thick
film. In this case the thickness $L_z$ of the volume, where the dissipation
occurs, will be determined by the thickness of the film, and according to
Eq.(\ref{biasfive}) this will give rise to a strong enhancement of the
friction.

For a planar film with thickness $d_f$ and dielectric constant $\epsilon _2$
on-top of a substrate with dielectric constant $\epsilon _3$, the reflection
coefficient is determined by
\begin{equation}
R_p=\frac{R_{p21}-R_{p23}\exp (-2qd_f)}{1-R_{p21}R_{p23}\exp (-2qd_f)}
\label{filmone}
\end{equation}
where
\begin{equation}
R_{pij}=\frac{\epsilon _i-\epsilon _j}{\epsilon _i+\epsilon _j},
\end{equation}
where index 1 is associated with vacuum. For a metallic film on a dielectric
substrate, or a metallic film on a metallic substrate with $\sigma _2\gg
\sigma _3,$ for $d_1\gg d_f$ and $R\gg d$ Eqs. (\ref{biastwo}) and (\ref
{filmone}) gives
\begin{equation}
\Gamma _f^c=\frac{w(V^2+V_0^2)R^{1/2}}{2^{9/2}\pi \sigma d_fd^{3/2}}.
\label{filmtwo}
\end{equation}
This is greater by a factor of $2d_1/d_f$ than the corresponding friction
for the infinitely thick sample. For thin film the effective resistivity of
the substrate is increased, giving rise to additional ohmic dissipation. In
\cite{Chumak} Eq.(\ref{filmtwo}) was obtained using a less general approach
and neglecting the spatial variation of the surface potential. The
conditions necessary for the validity of Eq.(\ref{filmtwo}) could not be
determined in this simplified approach.

\subsection{2D-system on-top of a dielectric or metal substrate.}

Let us now consider a 2D-system, e.g. electronic surface states or
a quantum well, or an incommensurate layer of ions adsorbed on a
metal surface. For example, for the Cs/Cu(100) system experiment
suggests the existence of an acoustic film mode even for the very
dilute phase ($\theta \approx 0.1$). This implies that the
Cs/Cu(100) adsorbate layer experience a negligible surface pinning
potential. The reflection coefficient for $p$-polarized
electromagnetic waves can be obtained using the approach proposed
in \cite {Langreth}. This gives (see \cite{Volokitin8}, detailed
derivation is given in Appendix A) :
\begin{equation}
R_p=\frac{1-1/\epsilon +4\pi qn_a\alpha _{\parallel }/\epsilon -qa(1-4\pi
n_aq\alpha _{\parallel })}{1+1/\epsilon +4\pi qn_a\alpha _{\parallel
}/\epsilon +qa(1+4\pi n_aq\alpha _{\parallel })},  \label{refcoefad}
\end{equation}
where $n_a$ is the concentration of the free carries of the charge per unit
area. The polarizability $\alpha _{\parallel }$ for the 2D-system in the
direction parallel to the surface is taken to be
\begin{equation}
\alpha _{\Vert }=-\frac{e^{*2}}{M(\omega ^2+i\omega \eta _{\Vert })},
\label{polarizability}
\end{equation}
where $\eta _{\Vert }$ is the damping constant, $e^{*}$ and $M$ are the
effective charge and the mass of the moving particles, respectively. In
comparison with the expression obtained in \cite{ Langreth}, Eq.(\ref
{refcoefad}) takes into account that the 2D-system is located a distance $a$
away from the image plane of the metal. Although this correction to the
reflection coefficient is of order $qa\ll 1$ , for a 2D-system on-top of a
good conductors ( $|\epsilon |\gg 1$), it gives the most important
contribution to the energy dissipation.

For good metals ($|\epsilon |\gg 1$), from Eq.(\ref{refcoefad}) we get
\begin{equation}
\mathrm{Im}R\approx \frac{2\omega \eta _{\Vert }qa\omega _q^2}{(\omega
^2-\omega _q^2)^2+\omega ^2\eta ^2},  \label{biassix1}
\end{equation}
where $\omega _q^2=4\pi n_ae^{*2}aq^2/M$. In the case of a 2D-structure
on-top of a dielectric, the factor $qa$ in Eq.(\ref{biassix1}) and in the
expression for $\omega _q^2$ must be replaced by $1/\epsilon ,$ where $%
\epsilon $ is the dielectric function of the substrate. Using (\ref{biassix1}%
) in (\ref{biastwo}) for $R\gg d$ we get
\begin{equation}
\Gamma _{ad}^c=\frac{w\eta MR^{1/2}(V^2+V_0^2)}{2^{9/2}d^{3/2}\pi n_ae^{*2}}.
\label{biasseven}
\end{equation}
This friction exhibits the same distance dependence as observed
experimentally \cite{Stipe}. The same expression for the friction is valid
for a 2D- structure on-top of a dielectric. Comparing Eqs.(\ref{biasthree})
and (\ref{biasseven}) we find that a 2D- structure on-top of a substrate
gives the same friction as for the clean surface with the effective
conductivity $\sigma _{eff}=n_ae^{*2}/M\eta 2d_1$. We obtain agreement with
experiment at $d=20$nm if $\sigma _{eff}\approx 4\cdot 10^9$s$^{-1}$. In the
case of a 2D-electron system, for $R=1\mu $m such an effective conductivity
is obtained if $\eta =10^{14}$s$^{-1}$ and $n_a=10^{15}$m$^{-2}$. For
Cs/Cu(100), for $n_a=10^{18}$m$^{-2}$ ($\theta \approx 0.1$) the electric
charge of the Cs ions $e^{*}=0.28e$ \cite{Senet}. Due to the similarities of
Cu and Au surfaces, a similar effective charge can be expected for the Cs/Au
surface. For such a 2D- system agreement with experiment is obtained for $%
n_a=10^{18}$m$^{-2}$ and $\eta =10^{11}$s$^{-1}$. In \cite{Volokitin8} we
estimated the damping parameter for a Cs atom associated with the covalent
bond $\eta _{\Vert cov}=3\cdot 10^9$s$^{-1}$ \cite{Volokitin8}. However the
collisions between the ions, and between the ions and other surface defects,
will also contribute to $\eta $. In this case $\eta _{col}\sim v_T/l$ where $%
v_T\sim \sqrt{k_BT/M}$, and $l$ is the ion mean free path. For $T=293$K and $%
l\sim 1$nm we get $\eta _{col}=10^{11}$s$^{-1}$.

For a spherical tip, with a 2D-system on-top of the substrate, from Eqs.(\ref
{biassix1}) and (\ref{biaseight}) for $R\gg d$ we get the contribution to
the friction from the 2D-system
\begin{equation}
\Gamma _{ad}^s=\frac{3RM\eta V^2}{2^6d\pi n_ae^{*2}}.  \label{biassix}
\end{equation}
At $d=20$nm this friction is $\sim $ two order of magnitude smaller than for
the cylindrical tip.

\subsection{Friction due to spatial fluctuations of static charge in the
bulk of the sample}

In this section we consider a dielectric substrate with a
stationary, inhomogeneous distribution of charged defects. Such a
situation was investigated experimentally \cite{Stipe} by
employing a fused silica sample irradiated with $\gamma $ rays. In
the course of irradiation, positively charged centers (Si dangling
bonds) are generated. Randomly distributed positive charges are
compensated by randomly distributed negative charges, thus on
average the sample is electrically neutral. We model the sample as
consisting of microscopically small volume elements $\Delta V_i$.
Each element is chosen sufficiently small that not more than one
charge center is present in it. The electric charge $q_i$ of each
element is equal to $\pm e$ or 0, in such away that the average
$\langle q_i\rangle =0$. We will consider the fluctuations of
charges in different volume element $i,j$ to be statistically
independent, so that $\langle q_iq_j\rangle =0$ for $i\neq j$. The
mean square of charge fluctuations within a given element $\langle
q_iq_i\rangle \approx 2ne^2$, where $n$ is the average number of
positive charges in one volume element. In the absence of the
cross terms the average tip-sample friction coefficient is
determined by adding friction coefficient from all charges $q_i$.
According to Eq.(\ref{biaseight}), the contribution to the
friction coefficient from charge $q_i$ in the element $ \Delta
V_i$ is given by
\begin{equation}
\Delta \Gamma _{i\Vert }=\lim_{\omega \to 0}ne^2\int_0^\infty dqq^2e^{-2qd_i}%
\frac{\mathrm{Im}R_p(\omega ,q)}\omega  \label{sf1}
\end{equation}
where $d_i=D(x_i,y_i)-z_i$. Here the coordinates $x_i,y_i,z_i$ give the
position of the $i$-th volume element in the substrate, and $D(x_i,y_i)$ is
the distance between the substrate and points $x_i,y_i$ located on the
surfaces of the tip. The total friction coefficient is obtained by summing
over all the elements. Replacing the sum by an integral ($n\sum \rightarrow
c\int d^3r$, where $c$ is the number of the positive charge centers per unit
volume), and integration over $z$ gives
\begin{equation}
\Gamma _{\Vert }=\lim_{\omega \to 0}\frac{ce^2}2\int_0^\infty dqq\int dx\int
dye^{-2qD(x,y)}\frac{\mathrm{Im}R_p(\omega ,q)}\omega  \label{sf2}
\end{equation}
For a cylindrical tip $D(x,y)=d+x^2/2R$, and we get

\begin{equation}
\Gamma _{\Vert }^c=\lim_{\omega \to 0}\frac{\sqrt{\pi R}ce^2w}2\int_0^\infty
dqq^{1/2}e^{-2qd}\frac{\mathrm{Im}R_p(\omega ,q)}\omega  \label{sf3}
\end{equation}

Using the same parameters as in Sec.2.2, for a gold tip separated by $d=10$%
nm from a dielectric sample with $c=7\cdot 10^{17}$cm$^{-3}$ we get $\Gamma
_{\Vert }=4.4\cdot 10^{-20}$kg s$^{-1}$.

For the tip surface with a 2D-structure on it, using Eq.(\ref{biassix1}) we
get
\begin{equation}
\Gamma _{2D\Vert }^c=\frac 1{2^{5/2}}\left( \frac e{e^{*}}\right) ^2\sqrt{%
\frac Rd}\frac{cw}{n_a}M\eta =\frac{e^2cw}{16\sigma _{eff}d}  \label{sf5}
\end{equation}
With $\sigma _{eff}=n_ae^{*2}/2M\eta d_1=4\cdot 10^9$, $c=7\cdot 10^{17}$cm$%
^{-3}$, and with the other parameters the same as before, we get for $d=10$%
nm, $\Gamma _{2D\Vert }^c=3.5\cdot 10^{-12}$kg s$^{-1}$, which is nearly the
same as was observed experimentally \cite{Stipe}. Thus our theory of
friction between a gold tip and silica substrate with an inhomogeneous
distribution of the charged defects is consistent with the theory of
friction between a gold tip and gold substrate ( see Section 2.4). In both
theories we have assumed that the gold surfaces are covered by a
2D-structure.

The study above has ignored the screening of the electric field in the
dielectric substrate. This can be justified in the case of very small tip-
sample separations (substantially smaller than screening length), as only
defects in the surface layer of thickness $d$ contribute to the integral in
Eq.(\ref{sf2}). When the screening is important, the effective electric
field outside the sample will be decreased by the factor $(\varepsilon +1)/2$
\cite{Landau1}, and the friction coefficient will be decreased by the factor
$((\varepsilon +1)/2)^2$, which is equal to $\approx $6.25 in the case of
silica. However. the inhomogeneity of the surface of the tip may be larger
than that of the sample surface, so that the damping parameter $\eta $ may
be larger for the 2D-structure on the surface of the tip. This increase in $%
\eta $ and screening effects will compensate each other.

\section{Van der Waals friction}

In this section we consider the van der Waals friction between two
surfaces covered by 2D-systems. The frictional stress between two
flat surfaces to linear order in the relative velocity $v$ can be
written in the form: $ \sigma =\gamma v$. According to
\cite{Volokitin6} in the case of the van der Waals friction the
contribution to the friction coefficient $\gamma _{\Vert } $ from
the $p$-polarized electromagnetic waves is given by
\[
\gamma _{\Vert }=\frac \hbar {2\pi ^2}\int_0^\infty d\omega \left( -\frac{%
\partial n}{\partial \omega }\right) \int_0^\infty dqq^3e^{-2qd}
\]
\begin{equation}
\times \mathrm{Im}R_{1p}\mathrm{Im}R_{2p}\frac 1{\left|
1-e^{-2qd}R_1pR_2p\right| ^2}  \label{vdwone}
\end{equation}
where $R_{1p}$ and $R_{2p}$ are the reflections coefficients for the
surfaces, and $n=[\exp (\hbar \omega /k_BT)-1]^{-1}$. In \cite
{Volokitin4,Volokitin5} we have shown that resonant photon tunneling between
two Cu(100) surfaces separated by $d=1$nm and covered by a low concentration
of potassium atoms gives rise to a friction six orders of the magnitude
larger than for clean surfaces. The adsorbate induced enhancement of the van
der Waals friction is even larger for Cs adsorption on Cu(100). In this
case, even at low Cs coverage ($\theta \sim 0.1$), the adsorbed layer
exhibit an acoustic branch for vibrations parallel to the surface \cite
{Senet}, and according to Eq.(\ref{refcoefad}), at small frequencies the
reflection coefficient is given by
\begin{equation}
R_p=1-\frac{2qa\omega _q^2}{\omega ^2-\omega _q^2+i\omega \eta }
\label{adsorbate1}
\end{equation}
where $\omega _q^2=4\pi n_ae^{*2}aq^2/M$. Using Eq. (\ref{adsorbate1}) in
Eq. (\ref{vdwone}) for
\[
\frac a{\eta d}\sqrt{\frac{4\pi n_ae^{*2}a}{Md^2}}\ll 1,
\]
gives
\begin{equation}
\gamma _{\Vert }\approx 0.62\frac{k_BTa^2}{\eta d^6}.  \label{adsorbate2}
\end{equation}
It is interesting to note that according to (\ref{adsorbate2}) $\gamma
_{\Vert }$ does not depend on $n_a$, $e^{*}$, and $M$. However, Eq.(\ref
{adsorbate1}) is only valid when there are acoustic vibrations in the
adsorbed layer. For Cs adsorbed on Cu(100) the acoustic vibrations exist
only for $\theta \ge 0.1$ \cite{Senet}. The friction coefficient for a
cylindrical atomic force microscope tip can be estimated using \cite
{Hartmann,Apell}
\begin{equation}
\Gamma _{\Vert }^c\approx 2w\int_0^\infty dx\gamma _{\Vert }(z(x))=0.68\frac{%
k_BTa^2R^{0.5}w}{\eta d^{5.5}}  \label{adsorbate3}
\end{equation}
where $R$ is the radius of the curvature of the tip and $w$ is its
width, and $\gamma _{\Vert }(z(x))$ the friction coefficient
between two flat surfaces at the separation $z(x)=d+x^2/2R$. In
Section 2 we have shown that the experimental data in \cite{Stipe}
can be explained by assuming that the gold surfaces are covered by
adsorbed layer of ions like Cs on Cu(100) with the damping
constant $\eta \approx 10^{11}$s$^{-1}$. With this value of $ \eta
$ and using $a=2.94${\AA } \cite{Senet}, $R=1\mu $m, $w=7\mu $m,
$T=293$ K we find that if $d<3$nm the contribution from the van
der Waals friction will dominate over the contribution from the
electrostatic friction. However, in the experiment a strong
enhancement in the friction was not observed at such short
separation. Thus, most likely a 2D-system of electronic origin is
responsible for the enhancement of the electrostatic friction. In
this case (see Section 2) $\eta _{el}\sim 10^{14}$s$^{-1}$ and the
van der Waals friction will give a negligible contribution for
practically all separations. Fig.2 shows how the friction between
the copper tip and the copper substrate depends on the distance
$d$ , when the surfaces of the tip and the substrate are covered
by a low concentration of the Cs atoms, and for clean surfaces. In
comparison, the friction between two clean surfaces at the
separation $d=1$nm is eleven orders of the magnitude smaller.
However, the friction between clean surfaces shown on Fig.2 was
calculated in the local optic approximation. For parallel relative
motion non-local optic effects are very important
\cite{Volokitin5}, and when it is taken into account, at $d=1$nm
the friction between adsorbate covered surfaces will be seven
orders of the magnitude larger than the friction between clean
surfaces.

\section{Phonon and internal non-contact friction}

\subsection{Non-contact friction due to excitation of substrate phonons}

Consider a tip which performs harmonic oscillation, $u=u_0\exp (-\mathrm{i}
\omega t)+c.c.,$ above an elastic body with a flat surface. This will
results in a fluctuating stress acting on the surface of the solid which
excite acoustic waves with parallel wave number $q<\omega /c_s,$ where $c_s$
is the sound velocity. The stress $\sigma_{iz}$ acting on the surface of the
elastic solid can be represented through the Fourier integral
\begin{equation}
\sigma_{iz}(\mathbf{x},t)=\int \frac{d^2q}{(2\pi)^2}\sigma_i(\mathbf{q}%
)u_0e^{i\mathbf{q} \mathbf{x}-i\omega t}+c.c.
\end{equation}
Using the theory of elasticity (assuming an isotropic elastic medium for
simplicity), one can calculate the displacement field $u_i$ on the surface $%
z=0$ in response to the surface stress distribution $\sigma_{iz}$
\begin{equation}
u_i(\mathbf{x},t)=\int \frac{d^2q}{(2\pi)^2}M_{ij}(\mathbf{q},\omega)
\sigma_j(\mathbf{q})u_0e^{i\mathbf{q} \mathbf{x}-i\omega t}+c.c.
\end{equation}
The energy dissipation per unit time equals
\[
P=\int d^2x \langle\stackrel{.}{u}_i(\mathbf{x},t) \sigma_{iz}((\mathbf{x}%
,t)\rangle=
\]
\begin{equation}
2\omega \int \frac{d^2q}{(2\pi)^2}\mathrm{Im}M_{ij}(\mathbf{q},\omega)
\sigma_i(\mathbf{q})\sigma_j^*(\mathbf{q})|u_0|^2  \label{int1}
\end{equation}
where $\langle...\rangle$ stands for the time averaging. The explicit form
of the stress tensor in the model of the elastic continuum is given in \cite
{ Persson3} (see also Appendix B). The energy dissipation per unit time must
be equal to $\Gamma <\stackrel{.}{u}(t)^2>=\Gamma 2\omega ^2\left|
u_0\right| ^2$. Comparing of this expression with (\ref{int1}) gives
\begin{equation}
\Gamma= \int \frac{d^2q}{(2\pi)^2}\frac{\mathrm{Im}M_{ij}(\mathbf{q},\omega)}
{\omega} \sigma_i(\mathbf{q})\sigma_j^*(\mathbf{q})  \label{phfr1}
\end{equation}

At typical experimental conditions we have $\omega \sim 10^3-10^6$s$^{-1}$
and $qr^{*}<\omega r^{*}/c_s$ $<10^{-3}<<1$, where effective radius of the
interaction $r^{*}\approx \sqrt{dR}$, and where $d$ is the separation
between the tip and the sample, and $R$ is the radius of curvature of the
tip. In Appendix B it was shown that in this case the contribution to the
friction from excitation of acoustic waves can be determined by calculating
the energy dissipation due to oscillating point force applied to the surface
of the semi-infinite elastic continuum. These calculations were done in the
connection with the vibrational energy relaxation of adsorbates \cite
{Persson1}. According to this theory the friction coefficient for vibration
of the tip normal to the surface is given by
\begin{equation}
\Gamma _{\perp }=\frac{\xi _{\perp }}{4\pi }\frac{K^2}{\rho c_t^3}
\label{phonon2}
\end{equation}
where $\xi _{\perp }\approx 1.65$, $c_t$ is the transverse sound velocity of
the solid, $\rho $ is the mass density of the sample, $K=\partial F/\partial
d,$ where $F(d)$ is the force acting on the tip due to interaction with the
sample.

In Appendix B it was shown that for vibration of the tip parallel to the
flat surface the friction coefficient due to excitation of the acoustic
waves is given by
\begin{equation}
\Gamma _{\Vert }=\frac{\xi _{\Vert }}{4\pi }\frac{\omega ^2}{\rho c_t^5}%
F_z^2(d)  \label{phonon3}
\end{equation}
where $\xi _{\Vert }\approx 1.50$. From the comparison of the Eqs.(\ref
{phonon2}- \ref{phonon3}) we get that $\Gamma _{\Vert }/\Gamma _{\perp }\sim
(\omega d/c_t)^2\ll 1$. We consider now two different contributions to the
tip-sample interaction.

\subsubsection{Van der Waals interaction}

Accordingly to the Lifshitz theory \cite{Lifshitz1} the stress $%
\sigma_{zz}(d)$ acting on the surface of two identical semi- infinite bodies
due to van der Waals interaction at small separation $d\ll c/\omega_p$
(where $\omega_p$ is the plasma frequency) and $d\ll \lambda_T$ is given by:
\begin{equation}
\sigma_{zz}(d)=\frac{\hbar}{8\pi^2d^3}\int_0^{\infty}d\xi\frac{%
[\varepsilon(i\xi)-1]^2} {[\varepsilon(i\xi)+1]^2}.  \label{ph1}
\end{equation}
In the Drude model the explicit form of $\varepsilon$ is
\begin{equation}
\varepsilon(i\xi)=1+\frac{\omega_p^2}{\xi(\xi+\eta)}  \label{ph2}
\end{equation}
For typical metal the damping constant $\eta\ll \omega_p$ and can be
neglected when integrating Eq.(\ref{ph1}). It follows from Eqs.(\ref{ph1})
and (\ref{ph2}) that
\begin{equation}
\sigma_{zz}=\frac{\hbar \omega_p}{32\sqrt{2}\pi d^3}  \label{ph3}
\end{equation}
For the spherical tip of radius $R$ using the same approximation
as in Eq.(\ref{adsorbate3}) we get
\begin{equation}
F_z(d)=\frac{R\hbar\omega_p}{32\sqrt{2}d^2}
\end{equation}
and
\begin{equation}
K^s=\frac{R\hbar\omega_p}{16\sqrt{2}d^3}
\end{equation}
Similarly, in the case of a cylindrical tip we have
\begin{equation}
F^c_z(d)=\frac{3wR^{1/2}\hbar\omega_p}{2^8d^{5/2}}
\end{equation}
and
\begin{equation}
K^c=\frac{15wR^{1/2}\hbar\omega_p}{2^9d^{7/2}}
\end{equation}
For copper tip separated from a copper substrate by $d=10$nm, and with $%
R=1\mu$m, $w=7\mu$m, we get for spherical tip $\Gamma_{\perp}^s=6.3%
\cdot10^{-18}$kg s$^{-1}$ and for cylindrical tip $\Gamma_{\perp}^c=1.3%
\cdot10^{-14}$kgs$^{-1}$. The phononic friction decreases as $d^{-6}$ and $%
d^{-7}$ for spherical and cylindrical tip, respectively.

\subsubsection{Electrostatic interaction due to a bias voltage}

In the presence of the bias voltage $V$ the attractive force between the tip
and the sample at $d\ll R$ is given by
\begin{equation}
F^c(d)=\frac{wV^2R^{1/2}}{2^{7/2}d^{3/2}}  \label{ph4}
\end{equation}
for a cylindrical tip, and
\begin{equation}
F^s(d)=\frac{RV^2}{4d}  \label{ph5}
\end{equation}
for a spherical tip. For bias voltage $V=1$Volt, and with the other
parameters the same as above, we get $\Gamma_{\perp}^s=8.8\cdot10^{-17}$kg s$%
^{-1}$ and $\Gamma_{\perp}^c=1.2\cdot10^{-13}$kgs$^{-1}$ for the spherical
and cylindrical tip, respectively. Note that in this case the friction
depends on the bias voltage as $V^4$.

For the vibrations of the tip parallel to the sample surface the expression
for the friction coefficient contains the addition small factor $(\omega
d/c_s)^2 \ll 1$. Thus the friction coefficient for parallel vibrations of
the tip will be by many orders of magnitude smaller than for normal
vibrations.

\subsection{Non-contact friction due to internal friction of the substrate}

In studying of the phononic friction in Section 4.1 it was assumed that the
deformations of the solids are purely elastic. However, the deformation will
be purely elastic or adiabatic only for infinitesimally small velocity, so
that at every moment of the time the system stays in the equilibrium state.
However, real motion always occurs with finite velocity, and the body does
not stay in equilibrium; and thus ``flow-processes'' occur, which tend to
bring it back to equilibrium. This leads to non-adiabatic deformations,
resulting in dissipation of the mechanical energy.

The energy dissipation is determined by two kind of processes. First, in the
presence of a temperature gradient in the body, result in heat flow.
Secondly, if in the body occurs some kind of internal motion, than
non-adiabatic processes occur, related with finite velocity of the motion;
these processes of energy dissipation can be denoted, as in liquids, as
internal friction or viscosity.

The friction coefficient due to the internal friction is determined by Eq.(%
\ref{phfr1}). However, in contrast to the phononic friction, large values of
$q\gg\omega/c_t$ play the most important role for the internal friction. For
$q\gg \omega/c_s$ the tensor component $M_{zz}$ is given by \cite{Persson3}

\begin{equation}
M_{zz}=\frac{2(1-\nu^2)}{Eq}  \label{intfr3}
\end{equation}
where $E(\omega)$ is the complex elastic modulus and $\nu$ is the Poisson
ratio.

\subsubsection{Van der Waals interaction}

For $R\gg d$ only the $\sigma _{zz}$ component of the stress tensor due to
the van der Waals interaction is important. In this case, for vibrations of
the cylindrical tip parallel to the sample surface, we get
\[
\sigma _z(\mathbf{q})=\int d^2xe^{i\mathbf{q}\mathbf{x}}\frac \partial
{\partial x}\sigma _{zz}(\mathbf{x})
\]
\begin{equation}
=-\frac{iq_xR^{1/2}}{2^7d^{5/2}}\frac{\sin (q_yw/2)}{q_y}(3+\xi ^2+3\xi
)e^{-\xi }  \label{intfr2}
\end{equation}
where $\xi =\sqrt{2dR}q_x$. Using (\ref{intfr2}) and (\ref{intfr3}) in (\ref
{phfr1}) we get for a cylindrical tip
\begin{equation}
\Gamma _{\Vert }^c=\frac{75\pi }{2^{16}}\frac{w\hbar ^2\omega _p^2}{d^6}%
\frac{\mathrm{Im}(E/(1-\nu ^2))}{\omega |E/(1-\nu ^2)|^2}  \label{int3}
\end{equation}
For the spherical tip similar calculations give
\begin{equation}
\Gamma _{\Vert }^s=\frac{0.25}{2^9\sqrt{2}\pi }\frac{R^{1/2}\hbar ^2\omega
_p^2}{d^{11/2}}\frac{\mathrm{Im}(E/(1-\nu ^2))}{\omega |E/(1-\nu ^2)|^2}.
\end{equation}
In general, $\mathrm{Im}[E(\omega )/(1-\nu ^2)]$ has many
resonance peaks, corresponding to different thermally activated
relaxation processes. One important source of internal friction at
high frequencies is related to thermal currents: elastic
compression of a material is commonly associated with heating
effects. If the compression takes place sufficiently rapidly,
there is no opportunity for heat to be conducted away, while for
very slow compression temperature gradients are eliminated by
thermal conduction. In both these cases the process of compression
will be reversible. In the former case it will be adiabatic and in
the latter one - isothermal. In both these limiting cases the
contribution from thermal current to the internal friction will be
negligible. However, in the intermediate frequency regime we
expect dissipation of mechanical energy into heat. The
characteristic frequency for the maximum dissipation will be of
order $\omega _t=1/\tau $ where, from dimensional arguments, we
expect the relaxation time $\tau \sim l^2/D,$ where $l$ is the
linear size of the compression region and $D$ the thermal
diffusibility $D=\kappa /\rho C_p$ (where $C_p$ is the specific
heat and $\kappa $ the heat conductivity). For $l\sim 10^3\AA$,
this gives for gold $\omega _t\approx 10^{11}s^{-1},$ which is
much higher than the resonance frequency of the cantilever of the
atomic force microscope. Another very important contribution to
the internal friction is point-defect flipping. This involves
thermally activated transitions of point defects or loose sites in
crystalline and amorphous network. A special case is the
vibrational motion of adsorbates on the surface of the substrate
and/or on the tip, as was treated separately above. Another
contribution to the internal friction comes from grain-boundary
slip \cite{McClintock}. For a
copper cylindrical tip and a copper substrate using $d=10$nm, $w=7\mu $m, $%
R=1\mu $m, $\omega =10^4$s$^{-1}$, and, as is typical for metals \cite
{Persson5}, $\mathrm{Im}E(\omega )/\left| E(\omega )\right| \approx 10^{-5}$
and $E\approx 10^{11}$N/m$^2$, gives $\Gamma _{\Vert }^c\approx 10^{-16}$kg$%
\cdot $s$^{-1}$. Thus at this separation the internal friction gives much
smaller contribution to the friction coefficient than electrostatic friction
due to bias voltage or spatial variation of the surface potential. However,
internal friction can give the dominant contribution for small separation $%
d\le 1$nm. For the spherical tip with $R=1\mu $m the friction coefficient is
two order of the magnitude smaller. Finally we note, as a curiosity, that
the internal friction of solids gives a very important contribution to the
rolling resistance of the most solids \cite{Persson6}, and is the main
contribution to rubber friction on rough substrates, e.g. road surface \cite
{Persson6}, where in the transition region between the rubbery and glassy
region of the rubber visco-elastic spectra, $\mathrm{Im}E(\omega )/\left|
E(\omega )\right| \approx 1$.

\section{Summary}

We have studied how the electrostatic friction between an atomic force
microscope tip and a substrate depends on: (a) the bias voltage, (b) the
spatial variation of the surface potential, and (c) the spatial fluctuation
of electric charge. We have found that the electrostatic friction can be
greatly enhanced in presence of a 2D-system on the surface of the sample or
on the tip. On metal surfaces such 2D-system can result from surface
electronic states, or from an incommensurate layer of adsorbed ions. We have
shown that the experimental data observed in \cite{Stipe} can be explained
by the electrostatic friction in presence of such a 2D-system. The theory
predicts the same magnitude, distance and bias voltage dependence of the
friction coefficient as it was observed in the experiment \cite{Stipe}, and
explains the bias- voltage-independent contribution to friction. The theory
of friction between a gold tip and silica substrate with an inhomogeneous
distribution of the charged defects is consistent with the theory of
friction between a gold tip and gold substrate. In both theories we have
assumed that the gold surfaces are covered by 2D-structure.

The electrostatic friction was compared with the van der Waals friction
arising from quantum and thermal fluctuations of the current densities
inside the bodies. The van der Waals friction as well as the electrostatic
friction can be greatly enhanced in presence of identical 2D-system on the
surfaces of the tip and the substrate. The van der Waals friction is
characterized by a stronger distance dependence than the electrostatic
friction, and may dominate at small separation. The van der Waals friction
between 2D-systems can be so large that it can be measured with present
state-of-the-art equipment.

Phonon and internal friction can be ruled out as mechanisms responsible for
non-contact friction observed in \cite{Stipe} because they predict stronger
distance and bias voltage dependence. For metal substrate the phonon
friction associated with excitation of acoustic phonons is negligibly small
in comparison with the electromagnetic friction (especially for motion of
the tip parallel to the substrate surface) because of small area in the
phase space available for these phonons.

\vskip 0.5cm \textbf{Acknowledgment }

A.I.V acknowledges financial support from Russian Foundation for
Basic Research (Grant N 06-02-16979) and ESF ``Nanotribology''.

\appendix

\section{Fresnel reflectivity for $p$-polarized electromagnetic waves with
2D-structure corrections}

We consider a semi-infinite metal having a flat surface which coincides with
the $xy$ plane, and with the $z$ axis pointed along the inward normal. The
metal surface is covered by an adsorbate layer located at $z=-a$. Let the $%
xz $ plane be the plane of incidence of evanescent electromagnetic plane
wave, with the parallel component of the wave vector $\mathbf{q}$ pointed
along the $x$-axis. The macroscopic electric field takes the form
\begin{equation}
\mathbf{E}=e^{iqx}\times \left\{
\begin{array}{r@{\quad;\quad}l}
\mathbf{I}e^{-pz}+\mathbf{R}e^{pz} & z<-a \\
\mathbf{A}e^{-pz}+\mathbf{B}e^{pz} & -a<z<0 \\
\mathbf{T}e^{-sz} & z>0
\end{array}
\right.
\end{equation}
where $p=(q^2-(\omega /c)^2)^{1/2}$, $s=(q^2-(\omega /c)^2\varepsilon
(\omega ))^{1/2}$, and $\varepsilon $ is the dielectric function of the
metal. According to \cite{Langreth} the boundary conditions at $z=-a$ can be
written in the form
\begin{eqnarray}
A_ze^{pa}+B_z^{-pa}-R_ze^{-pa}-I_ze^{pa} &=&4\pi pn_a\alpha _{\Vert }\left(
R_ze^{-pa}-I_ze^{pa}\right)  \label{B1} \\
B_ze^{-pa}-A_ze^{pa}-R_ze^{-pa}+I_ze^{pa} &=&-\frac{4\pi qn_a\alpha _{\perp }%
}p\left( R_ze^{-pa}+I_ze^{pa}\right)  \label{B2}
\end{eqnarray}
where $\alpha _{\Vert (\perp )}$ is the polarizability of the adsorbate in
the direction parallel (normal) to the surface. From the ordinary boundary
conditions at $z=0$ it follow
\begin{equation}
B_z=\frac{\varepsilon p-s}{\varepsilon p+s}A_z  \label{B3}
\end{equation}
For a 2D-system $\alpha _{\perp }=0,$ and for $q\gg \omega /c$ and $qa\ll 1$
Eqs.(\ref{B1}-\ref{B3}) give the reflection coefficient $R_z$, given by Eq.(
\ref{refcoefad}).

\section{Friction coefficient due to excitation of the acoustic waves}

According to \cite{Persson3} the tensor $\stackrel{\leftrightarrow }{\mathbf{%
\ M}}$ in Eq.(\ref{int1}) is given by
\[
\stackrel{\leftrightarrow }{\mathbf{M}}=\frac i{\rho c_t}\Big
(\frac 1{S(q,\omega }\Big[Q(q,\omega )(\hat z\mathbf{q}-\mathbf{q}\hat z)
\]
\begin{equation}
+\left( \frac \omega {c_t}\right) ^2(p_l\hat z\hat z+p_t\hat q\hat q)\Big] +%
\mathbf{nn}\frac 1{p_t}\Big)  \label{apGone}
\end{equation}
where $\hat q=\mathbf{q}/q$, $\mathbf{n}=\hat z\times \hat q$, and where
\begin{eqnarray}
S &=&\left( \frac{\omega ^2}{c_t^2}-2q^2\right) ^2+4q^2p_tp_l, \\
Q &=&2q^2-\omega ^2/c_t^2+2p_tp_l, \\
p_t &=&\sqrt{\frac{\omega ^2}{c_t^2}-q^2},\,\,\,p_l=\sqrt{\frac{\omega ^2}{%
c_l^2}-q^2}
\end{eqnarray}
In the equations above, $\rho $, $c_t$, and $c_l$ are the mass density and
the transverse and longitudinal sound velocities of the solid, respectively.
Note that $c_t$ and $c_l$ are in general complex frequency dependent
quantities given by
\begin{eqnarray}
c_t^2 &=&\frac E{2\rho (1+\nu )}, \\
c_l^2 &=&\frac{E(1-\nu )}{\rho (1+\nu )(1-2\nu )}
\end{eqnarray}
where $E(\omega )$ is the complex elastic modulus and $\nu $ is the Poisson
ration.

The acoustic waves have wave number $q<\omega /c_t$. At typical experimental
condition the frequency of the vibrations of the tip $\omega \sim 10^3-10^6$
s$^{-1}$ and $qR_{int}<\omega R_{int}/c_t\ll 1$, where $R_{int}\sim \sqrt{dR}
$ is the radius of the interaction of the tip with the sample surface. In
this case for the vibrations of the tip normal to the surface we get
\[
\sigma _{\perp i}(q)=\int d^2xe^{i\mathbf{q}\mathbf{x}}\frac \partial
{\partial d}\sigma _{iz}^0(\mathbf{x},d)\approx
\]
\begin{equation}
\delta _{iz}\int d^2x\frac \partial {\partial d}\sigma _{zz}^0(\mathbf{x}%
,d)=\frac \partial {\partial d}F_z(d)  \label{apGthree}
\end{equation}
where $\sigma _{iz}^0$ is the static stress acting on the surface of the
sample. Using Eqs.(\ref{apGthree}) and (\ref{apGone}) in Eq.(\ref{phfr1}) we
get
\begin{equation}
\Gamma _{\perp }=\frac{\xi _{\perp }}{4\pi }\frac{K^2}{\rho c_t^3}
\end{equation}
where $\xi _{\perp }=\xi _{\perp l}+\xi _{\perp t}+\xi _{\perp s}$, $%
K=\partial F_z/\partial d$ and where the contributions from the longitudinal
$\xi _{\perp l}$, the transverse $\xi _{\perp t}$, and surface (Rayleigh) $%
\xi _{\perp s}$ acoustic waves are given by
\begin{eqnarray}
\xi _{\perp l} &=&\int_0^{c_t/c_l}dx\frac{\sqrt{(c_t/c_l)^2-x}}{(1-2x)^2+4x%
\sqrt{(1-x)}\sqrt{(c_t/c_l)^2-x}},  \label{apGfour} \\
\xi _{\perp t} &=&\int_{c_t/c_l}^1dx\frac{4x[x-(c_t/c_l)^2]\sqrt{1-x}}{%
(1-2x)^4+16x^2[x-(c_t/c_l)^2](1-x)},  \label{apGfive} \\
\xi _{\perp s} &=&\pi \sqrt{x_c-(c_t/c_l)^2}/f^{\prime }(x_c),
\label{apGsix}
\end{eqnarray}
where
\begin{equation}
f(x)=4x\sqrt{x-1}\sqrt{x-(c_t/c_l)^2}-(2x-1)^2,
\end{equation}
and where $x_c$ is the solution of the equation $f(x)=0$, $f^{\prime
}=df(x)/dx$. In Eqs.(\ref{apGfour}-\ref{apGsix}) the sound velocities $c_t$
and $c_l$ are assumed real, taken at $\omega =0$.

For the vibrations of the tip parallel to the surface the main contribution
to the energy dissipation due to excitation of the acoustic waves gives
component of $\sigma _i$ which acts in the $z$-direction. For this component
we get
\begin{equation}
\sigma _{\Vert z}(q)=\int d^2xe^{i\mathbf{q}\cdot \mathbf{x}}\frac \partial
{\partial x}\sigma _{zz}^0(\mathbf{x})\approx iq_xF_z(d)  \label{apGseven}
\end{equation}
Using Eqs.(\ref{apGseven}) and (\ref{apGone}) in Eq.(\ref{phfr1}) we get
\begin{equation}
\Gamma _{\Vert }=\frac{\xi _{\Vert }}{8\pi }\frac{\omega ^2}{\rho c_t^5}%
F_z^2(d)
\end{equation}
where $\xi _{\Vert }=\xi _{\Vert l}+\xi _{\Vert t}+\xi _{\Vert s}$,
\begin{eqnarray}
\xi _{\Vert l} &=&\int_0^{c_t/c_l}dxx\frac{\sqrt{(c_t/c_l)^2-x}}{(1-2x)^2+4x%
\sqrt{(1-x)}\sqrt{(c_t/c_l)^2-x}},  \label{apGfour} \\
\xi _{\perp t} &=&\int_{c_t/c_l}^1dxx\frac{4x[x-(c_t/c_l)^2]\sqrt{1-x}}{%
(1-2x)^4+16x^2[x-(c_t/c_l)^2](1-x)},  \label{apGfive} \\
\xi _{\perp s} &=&\pi x_c\sqrt{x_c-(c_t/c_l)^2}/f^{\prime }(x_c),
\label{apGsix}
\end{eqnarray}
For most metals $c_t/c_l\approx 1/2$ and for this case $\xi _{\perp }=1.62$
and $\xi _{\Vert }=1.50$.

\vskip 0.5cm FIGURE CAPTIONS

Fig. 1. Scheme of the tip-sample system. The tip shape is characterized by
its length $L$ and the tip radius of curvature $R$.

Fig.2. The friction coefficient associated with the van der Waals friction
between a copper tip and a copper substrate, both covered by low
concentration of cesium atoms, as a function of the separation $d$. The
cylindrical tip is characterized by the radius of curvature $R=1\mu $m and
the width $w=7\mu $m. The other parameters correspond to Cs adsorbed on
Cu(100) at the concentration $n_a=10^{18}$m$^{-2}$ (coverage $\theta \approx
0.1$) \cite{Volokitin8,Senet}: $e^{*}=0.28e,$ $\eta =10^{11}$s$^{-1}$, $%
a=2.94$\AA , $T=293$K. (The base of the logarithm is 10)

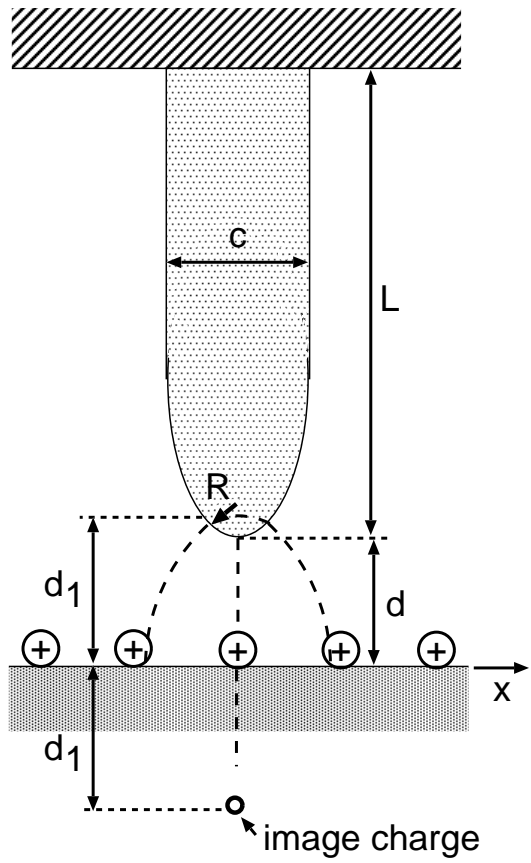
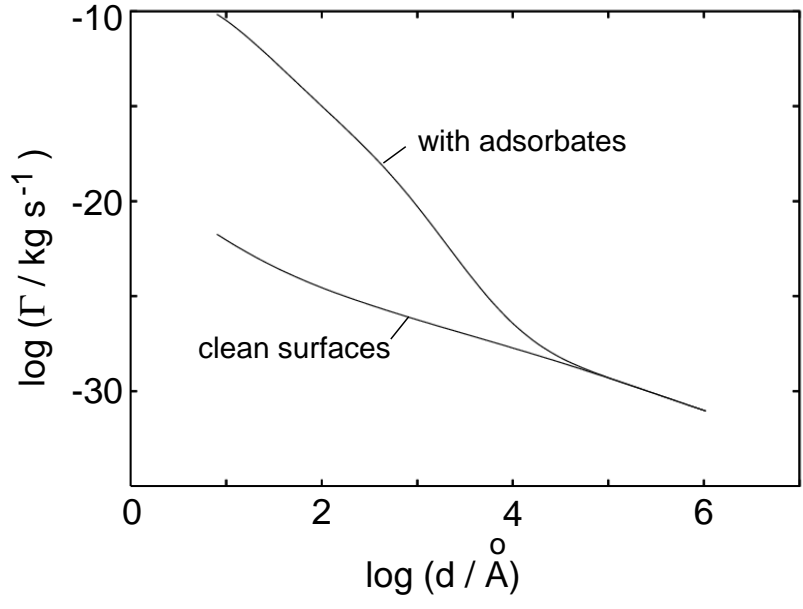
\end{document}